# Direct assessment of individual connotation and experience
*An introduction to cognitive-affective mapping*

**Lisa Reuter** , *Albert Ludwigs University Freiburg*

**Jordan Mansell**, *Network for Economic and Social Trends, Western University*

**Carter Rhea**, *L'Université de Montréal*

**Andrea Kiesel**, *Albert Ludwigs University Freiburg*

ABSTRACT. We introduce cognitive-affective maps (CAMs) as a novel tool to assess individual experiences and belief systems. CAMs were first presented by the cognitive scientist and philosopher Paul Thagard as a graphical representation of a mental network, visualizing attitudes, thoughts, and affective connotations toward the topic of interest. While CAMs were originally used primarily to visualize existing data, the recent release of the new software tool *Valence* has facilitated the applicability of CAMs for empirical data collection. In this article, we explain the concept and the theoretical background of CAMs. We exemplify how CAMs can be applied in research practice, including different options for analysis. We propose CAMs as a user-friendly and versatile methodological bridge between qualitative and quantitative research approaches and encourage incorporating the method into studies to access and visualize human attitudes and experience.

We introduce cognitive-affective mapping as a novel, innovative method for data collection that can bridge gaps between quantitative and qualitative research methods. A cognitive-affective map (CAM) is a graphical representation of a network, visualizing attitudes and thoughts toward the topic of interest. Unlike other network representations, such as mind maps or directed acyclic graphs, CAMs capture not only the interconnectedness of cognitive concepts but also their affective valences. The method was developed and introduced by the philosopher Paul Thagard (2010) and used primarily in conflict research as a tool to depict and communicate divergent perspectives. In this regard, CAMs might be considered an extension of cognitive maps (e.g., Boukes et al., 2020) that allow the integration as well as the differentiation of cognitive concepts (Conway et al., 2014; Neumann, 1981; Suedfeld, 2010; Tetlock, 1983). While these cognitive maps are able to depict the "cognitive complexity" of a given topic, CAMs additionally allow one to depict the affective valences—that is, whether a concept is considered as positive, negative, neutral, or ambivalent. As we discuss later, the inclusion of affective valences makes it possible to calculate a number of cognitively meaningful network properties.

Recently, Rhea et al. (2020) developed *Valence*, an open-source browser-based drawing tool for CAMs that records each graph's network properties. This dual innovation in practicality and quantifiability allows CAMs to be used for large-scale empirical research. Through the *Valence* application, CAMs can be created in field or lab settings with research assistance available and in online studies after standardized instructions. Furthermore, *Valence*'s data exporting allows for qualitative, quantitative, and mixed-method data analyses. In this article, we first provide a detailed explanation of the concept of CAMs (1), followed by the method's theoretical background (2). Then we exemplify CAMs' utility as a research tool with an example (3) before discussing export and analysis possibilities of CAMs' network properties (4). We conclude by discussing the larger implications of CAMs for scientific research progress, particularly the opportunities to bridge gaps between qualitative and quantitative research (5), and we present methodological application examples (6). We end with a summary of our CAM presentation (7).

## 1. What is a CAM?

A CAM is visualized as a network consisting of nodes (vertices; concepts) and edges (links between the nodes). Figure 1 summarizes the properties of a CAM. The nodes can represent any content in text form, for example, thoughts, events, emotions, or factual knowledge. Each node also conveys an affective valence, represented by the color and the shape of the node. There are four different colors and shapes: green ovals represent positive valence; red rectangles represent negative valence; neutral is represented by yellow rectangles, meaning that the node is linked to neither positive nor negative affect; ambivalent is indicated by a superimposed oval and rectangle in purple and means that the content of the node is emotionally ambivalent—in other words, there are simultaneously positive and negative feelings. There are three valence levels to choose from for both positive and negative valence; the thicker the border, the more intense the affective connotation. The neutral and ambivalent evaluations are without intensity levels.

The network's nodes can be linked via edges. Edges are connecting lines that are either solid or dashed. Solid lines imply that the two nodes are positively correlated (or mutually reinforcing concepts), whereas dashed lines imply a negative correlation (or concepts in conceptual opposition). In the original CAM model developed by Thagard (2010), edges were undirected and did not imply causality. In more recent studies, solid and dashed arrows were (optionally) added to imply causal directionality allowing CAMs to be analyzed as directed Markov graphs.

The combination of qualitative information captured by conceptual nodes and quantitative information captured by network properties makes CAMs a unique tool to study individual connotation and experience. While other mapping approaches incorporate both conceptual content and quantifiable network properties, such as fuzzy cognitive maps (Kosko, 1986), the CAM method can be distinguished for its applicability to study individual-level differences.

## 2. Theoretical development

Thagard (2010) introduced the concept of CAMs in his workshop paper "*EMPATHICA*: A computer support system with visual representations for cognitive-affective mapping." The method is based on his theories about the relationship between individuals' perspectives and their network's structure and valences. In his book *Coherence in Thought and Action*, Thagard (2000) combined philosophy and cognitive science to explain how people unify their thoughts and experiences coherently. He also referred to constraint satisfaction in networks of connected representations that coherently fit together (positive constraints) or incoherently do not fit together (negative constraints). In 2006, Thagard revised and expanded his coherence theory in his book *Hot Thought*. He elaborated on the emotional component of human thought to explain how cognition and emotion interact in everyday human thinking. In his extended HOTCO (hot cognition) model, Thagard (2006) included emotional coherence and assigned a valence value to the network elements.

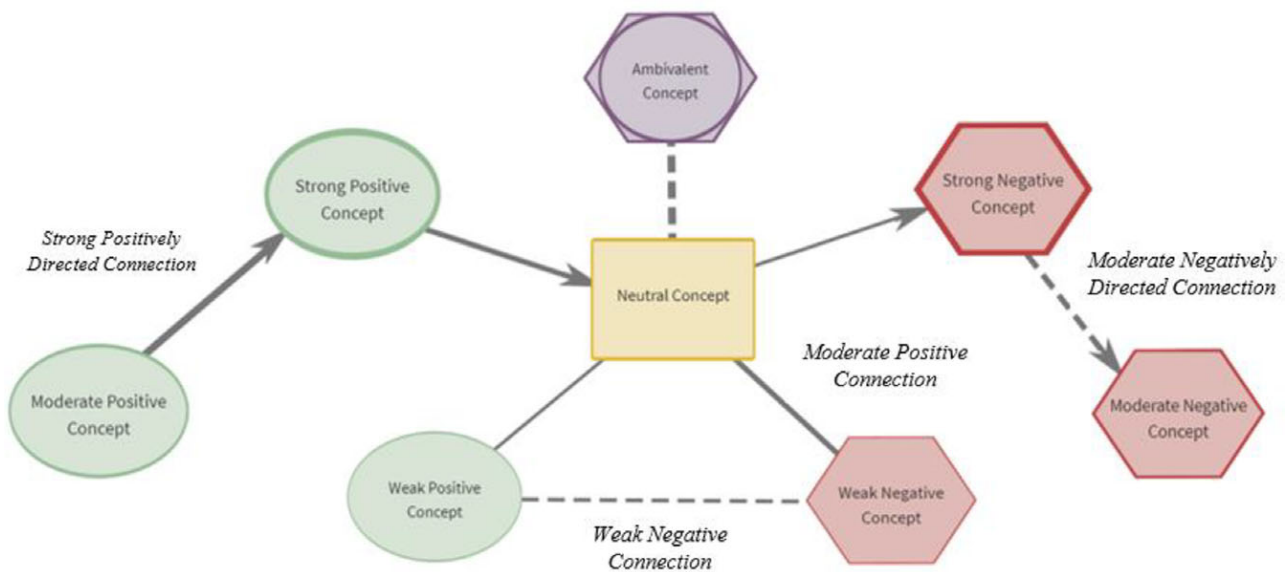

**Figure 1.** Summary of CAM properties.

Synthesizing these ideas, Thagard (2010) presented the software application *EMPATHICA*, with which CAMs can be created, along with an example of how CAMs can be used for conflict resolution by increasing mutual understanding. Using *EMPATHICA*, two different parties can create visual representations of their perspectives that can be qualitatively compared to identify points of conflict. In the study of conflict reconciliation, several studies have applied the CAM approach to visualize contrary positions and derive possible solutions for stakeholders (Homer-Dixon et al., 2014; Thagard, 2015a) or to graphically represent attitudes and changes in attitudes in response to interventions (Thagard, 2012a, 2018; Wolfe, 2012). As a secondary application, CAMs were often used to visualize data obtained in interviews or text analysis; for example, Luthardt et al. (2020) used CAMs in their research on social innovation for early childhood education, while Homer-Dixon et al. (2013) used CAMs to map conceptual structures in ideological research.

A limitation of past applications of CAMs was that in most instances, researchers drew CAMs themselves based on a critical assessment of collected data, such as interviews or written text. This research-centered approach suffers from two limitations: (1) the development of a single CAM based on a critical assessment of collected data is time-consuming, and (2) the reliance on critical assessment introduces bias from subjective (author) assessment into the results. To overcome these limitations, the previous *EMPATHICA* software was modified into a new version called *Valence*, which has a freely available code (Rhea et al., 2020). Unlike *EMPATHICA*, *Valence* records the network statistical properties of CAMs providing researchers with a mechanism to objectively study and compare individual CAMs. A further feature of the *Valence* update is the development of a web-based browser interface, which instructs participants to draw CAMs themselves while allowing data collection in both an online and offline environment. In the proceeding section, we demonstrate the applicability of CAMs using a visual example.

## 3. Applicability and example

CAMs can capture complex attitudinal interrelationships to gain essential insights into human decision criteria and motivational structures. Because of the adaptability of its content and design, the method can be used in diverse disciplines and can contribute to interdisciplinarity. To date, while CAMs have been used primarily in a practical sense to access or represent cognition, the method is applicable to scientific research—for example, CAMs can be used to study variations in individual motivation and decision processes or the interplay between cognition and emotion. Therefore, the method can be used in many different contexts, for instance, in psychiatric treatments or assessments. For example, in treating psychological disorders CAMs can be used to map emotional change in psychotherapy (Thagard & Larocque, 2020).

Here we demonstrate the applicability of CAMs using an example taken from a CAM study about individuals' experience with the coronavirus pandemic (Mansell, Reuter, et al., 2021). Participants were first instructed how cognitive-affective mapping works and how they can draw a CAM with *Valence*. Note that attention checks can be used to improve response validity. Figure 2 shows sample slide excerpts from the instruction set for the *Valence* application using a neutral example topic, while the complete sample instruction is available in the supplementary materials. Then participants were instructed to draw a CAM on the topic of the coronavirus pandemic. Figure 3 shows the CAM of one participant as an example.

In summary, the vivid appearance and transparent rules of CAMs make them accessible to laypersons. As most people have some familiarity with mind mapping methods like directed acyclic graphs, word bubbles, or decision trees, the approach of cognitive-affective mapping is perceived as familiar and agreeable. Also, CAMs benefit from their high degree of face validity: what is represented by a CAM, including the interconnections between concepts, is immediately recognizable for most participants. In our experience, this transparent character increases the acceptance and willingness to participate in CAM studies. Furthermore, their latent and valence network properties make them applicable for research on cognition, emotion, and experience, or studies on individual beliefs, perceptions, and evaluations. Research using CAMs may be further supplemented with experimental or longitudinal (repeated) research designs.

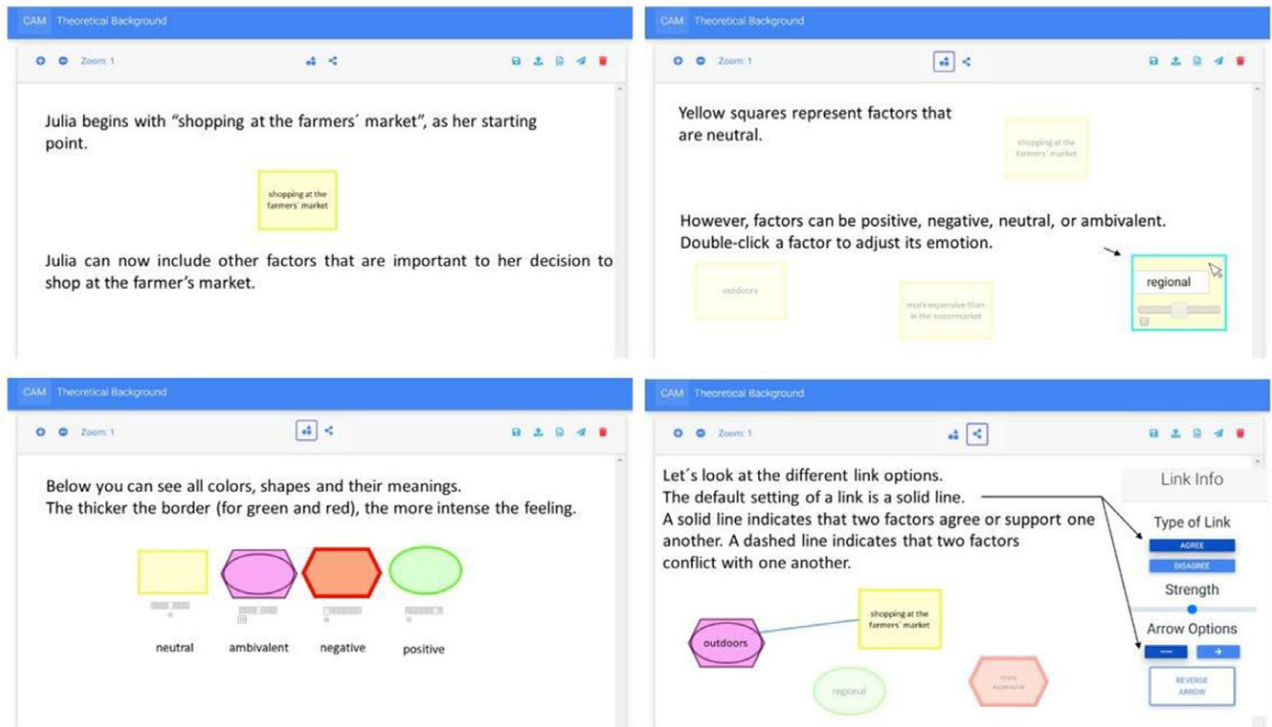

**Figure 2.** Exemplary instruction slides for drawing a CAM with the *Valence* application. Note that these are only excerpts. For detailed instructions, please see the supplementary materials.

The analysis of connotation and experience using CAMs using network properties is a recent methodological development; therefore, further validation is required to ensure that the interpretation of CAMs by researchers are consistent with the interpretation by participants. As an initial assessment of CAMs' validity, Mansell, Mock, et al. (2021) calculated the probability of randomly replicating each CAM obtained in their study using a Bayesian inference algorithm. Their results provide evidence that the CAMs in their study are a result of intentional decisions by participants and not random imputation.

Previous CAM studies raised fundamental questions regarding CAM analyses and interpretations that need to be clarified in future studies. In particular, there is the question of how ambivalent nodes are interpreted and coded for further calculations. For example, Mansell, Mock, et al. (2021) used two different codings for ambivalent nodes when calculating the affective network parameters: one time they treated ambivalent like neutral nodes (coded 0), and one time they omitted ambivalent nodes (and only included positive and negative nodes). Another issue comes with the interpretation of the connecting links. The two qualitatively different kind of links, supportive and inhibitory, do provide a more accurate representation of the nature of the connections; however, there is currently a lack of research on the extent to which participants understand the distinction between the connections in the way intended by the researcher. Specifically, establishing an understanding of inhibitory connections proves challenging. To ensure the validity of CAMs as a research tool, future studies will need to investigate these issues.

## 4. Export and analysis of network properties

The content of any single CAM can be exported as a set of simple network properties, node values, and edge values, allowing the data to be analyzed using graphical analysis. The data can be downloaded as tabular files (.csv). Image files (.png), suitable for qualitative analysis, can be exported for each CAM and include a participant ID that can be matched to the tabular files. The tabular data include both latent (structural) and valence (emotional/nodal) properties of CAMs. This includes the affective valence and strength of each node and information about which nodes are connected, and the strength and potential directionality of these connections. While the x- and y-positions of each node in the graphical space are recorded, other calculations such as the physical distance between nodes (e.g., 1 cm) are not recorded, as this information is not relevant for network analysis. Using this tabular information, a variety of network properties can be calculated to analyze and compare the properties of CAMs, including measures of affective content such as the average valence of a network, the dependency of the network to a specific emotion, the diversity of emotional properties and connections (Simpson's diversity index), or structural measures such as centrality, diversity, and transitivity. For an overview of CAMs' network parameters, see Table 1.

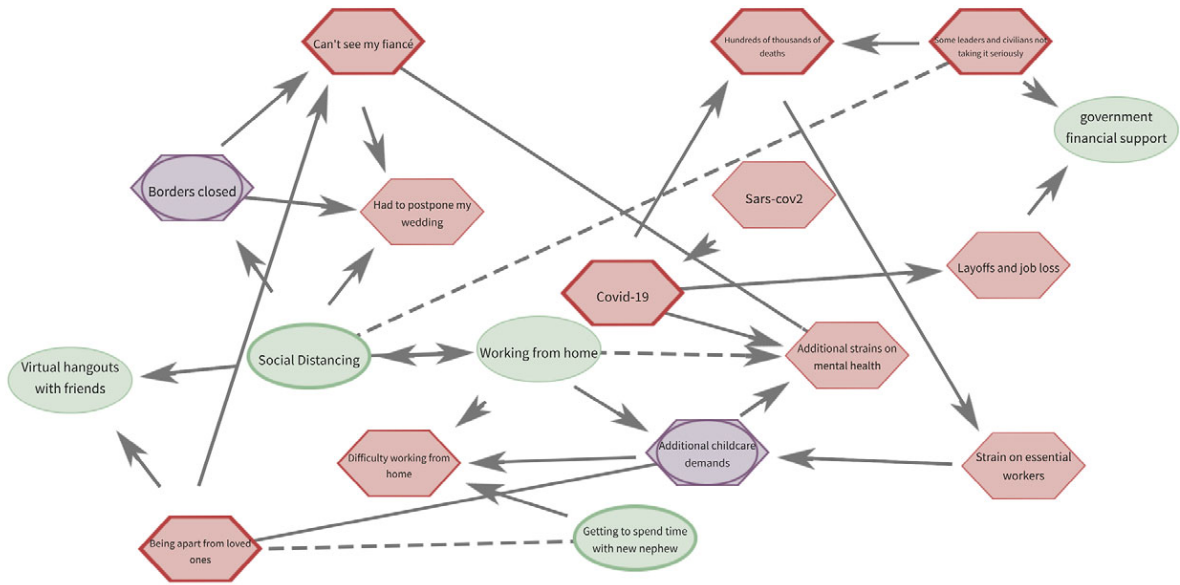

**Figure 3.** Exemplary CAM capturing the experience of one participant with the coronavirus pandemic. The academic instruction was as follows: "We are interested in capturing your experience, the events, thoughts, and feelings resulting from the current coronavirus outbreak. Using the mapping tool, please draw everything that comes to mind concerning your experience with the coronavirus. Think about what matters in the current coronavirus outbreak, and please do your best to draw everything that comes to your mind concerning the coronavirus" (Mansell, Reuter, et al., 2021).

Currently, two versions of the *Valence* software are freely available to access online. One is operated by the Cascade Institute, the other by the University of

**Table 1.** Exemplary network parameters of a CAM.

| Measure | Description | Scale |
|---|---|---|
| Average Valence | Mean value of all node valences of a CAM.[a] | -3-3 |
| Valence Percentages | Percentages of the individual valence options (e.g., for positive, negative, neutral or ambivalent). | 0-1 |
| Central Node Valence | Valence of the most central node.[a] | -3-3 |
| Centrality | The number of links on a node normalized by the total number of possible links. | 0-1 |
| Density | The number of a CAM's links, divided by the total number of possible links. | 0-1 |
| Diameter | Maximum distance from one node to another. | |
| Number of Nodes | Total number of nodes. | |
| Number of Links | Total number of links (measure can be subdivided in number of supporting/contradicting links/arrows). | |
| Triadic Closure/Transitivity | Total number of triangles (three nodes connected with each other by links) divided by total number of possible triangles. | 0-1 |
| Simpson's Diversity | The extent to which a network employs heterogeneous properties. | 0-1 |

[a] Neutral nodes are usually counted as zero, while ambivalent nodes could be counted as zero (one value) or as −1,5 +1,5 (zero in sum, but two values).

Freiburg. The University of Freiburg software version supports arrows (directedness) as graphical features. The software code for *Valence* is freely available (Rhea et al., 2020). A code for analyzing the original CAM data sets is also freely available. With this code, it is easily possible to clean and combine data sets and calculate the aforementioned network parameters, such as density. An interactive website, with which such calculations can be automated in a user-friendly way, is in progress.

## 5. Bridging gaps in qualitative and quantitative research

Nowadays, both qualitative and quantitative research methods are recognized as significant to scientific research, however, after many paradigmatic controversies about scientific advancement, gaps still exist between qualitative and quantitative research. Recent developments in mixed-methods designs have bridged many of these gaps, yet some are more enduring. While structured interviews can provide detailed accounts of individual experiences, it is difficult to directly contrast the substance of these accounts in an objective and quantifiable manner. Instead, researchers must insert themselves into the discussion to provide meaningful context at the cost of risking subjective interpretive bias. Also problematic, survey responses are highly constrained and can be influenced by factors such as question order, wording, and scale (see Tourangeau et al., 2000).

When incorporated into a mixed-method design, CAMs contain a number of properties that can assist researchers in reducing these biases. In contrast with interviews, CAMs' properties allow unbiased comparisons of quantitative and qualitative data between subjects. Compared with closed-response questions, CAMs allow participants to express themselves openly without the constraints of instrument bias. In this regard, CAMs mimic interviews but also capture quantifiable information. Additionally, for large-scale projects, CAMs are faster to implement than structured interviews as they can be administered to multiple participants at a time. A further benefit of CAMs' accessibility, or "face value," is that CAMs' content is easy to visualize and useful to facilitate communication both during and after data collection.

The simplicity of the CAM method can also be viewed critically, considering the reduction of emotions to mere affective connotation—that is, to the four categories *positive*, *negative*, *neutral*, and *ambivalent*. Various disciplines, including philosophy and neuroscience in addition to psychology, have come up with theories and models of emotions. Prominent are two-dimensional (dual process) models that emphasize valence/(un)pleasantness and arousal/(de)activating as basic dimensions of emotion, also named core affect (Russell, 2003). However, other researchers propose additional and different dimensions (e.g., Fontaine et al., 2007; Scherer et al., 2006). Hence, the CAM categorization entails a loss of information regarding the specifics and nuances of emotions. Thus, researchers using the CAM method should be aware that CAMs cannot be a decided explanation and differentiated consideration of discrete and specific emotions—instead, they offer the possibility to incorporate simple and fundamental valence information, to identify cognitive and emotional coherence and to strengthen interdisciplinary collaboration. We emphasize for readers that we are *not* advocating the abolition of structured interviews or survey response measures; rather, we argue that CAMs can be incorporated into studies to provide an objective comparative measure that was not previously available.

## 6. Examples for specific research designs

In this section, we point out different methodological approaches for the application of CAMs: They can be used to visualize existing information (6.1.), be integrated into experimental studies to measure the effect of interventions on attitudes and beliefs (6.2.), collect and analyze individual-level data (6.3).

### 6.1. Experts draw CAMs to visualize pre-collected data

The original primary application of the CAM method was conflict resolution (Thagard, 2010). Findlay and Thagard (2014) presented the Camp David negotiations in 1978 as a specific example. They used CAMs to visualize the change of attitudes of the two negotiators. The descriptions in Jimmy Carter's memoirs served as the basis for the mapping. Findlay and Thagard drew multiple CAMs for both conflict parties to depict the emotional changes that were central to conflict resolution over the course of the negotiation.

Other studies have used CAMs to represent disparate worldviews of different cultures (Thagard, 2012a); to signify values in scientific thinking (Thagard, 2012b); to illustrate social conflict (Homer-Dixon et al., 2013); to capture the cognitive-affective structures of ideologies (Homer-Dixon et al., 2014), ethical conflicts (Thagard, 2015b), or political beliefs (Thagard, 2018); as well as to represent analogical and emotional aspects of allegory structures (Thagard, 2011). Thagard and Larocque (2020) also suggested that CAMs can be used in psychotherapy to help patients better understand and positively change their emotional states. Wolfe et al. (2012) suggested using CAMs to mediate between stakeholders in water policy, and Luthardt et al. (2020) used CAMs to visualize perceptions and values related to innovation transfers in early childhood education as part of a triangulation approach.

### 6.2. Using CAMs for experimental and correlation studies

CAMs can be used in quasi-experimental and experimental designs to investigate cause-effect correlations. Here, the focus of investigation is the change in structural properties between treatment and control groups. Because individual CAMs have the potential to vary significantly in both emotional and latent structures, researchers may wish to consider implementing a within-subjects design. For example, Reuter et al. (2021) had participants draw a CAM on the topic of the coronavirus pandemic. After completing this task, participants in the experimental group were instructed to go for a one-hour walk, while the control group pursued any activity at home for one hour. Afterward, all participants were asked to draw a second CAM on the same topic. Differences in pre/post-CAM change were quantitatively examined through overall CAM valence, valence of the central coronavirus pandemic concept, and other structural measures such as the number of positive/negative nodes. The results showed substantive differences between the control and treatment group that appear to be related to higher levels of introspection and reflections among the treatment group. On the basis of these differences, CAMs may be useful to study how different activities influence attitude formation and change.

CAMs are also highly applicable for survey research. A classic cross-sectional design considers the correlation between two or more characteristics from a single sample data collection. In addition to correlations within the network (between specific network parameters, e.g., between *density* and *number of nodes*), external variables, such as questionnaire data, can also be collected and correlated with network structures. To test correlations inferentially, classical significance tests, especially correlation and regression analysis, can be used. Mansell, Reuter, et al. (2021) collected CAMs on the topic of the coronavirus pandemic and, at the same time, questionnaires on attitudes toward the pandemic. A central questionnaire scale was the perceived threat of the coronavirus. They examined the extent to which this score could be predicted by structural features of the network. In a second independent study, Mansell, Mock, et al. (2021) examined the relationship between CAMs' emotional and network structures and the introduction of the federal Canadian carbon tax. Both studies show significant correlations between participants' attitudes and CAMs' network structures.

Furthermore, the *Valence* software allows for flexible project planning in that participants can be presented with a list of predefined concepts, which can be arranged into a CAM (as in Reuter et al., 2021) or asked to generate and relate their own unique set of concepts (as in Mansell, Reuter, et al. 2021).

### 6.3. Analyzing the content of CAMs

The previous example studies focused on structural network parameters and took less account of the content of the different nodes of the CAMs. Various approaches to capture the nodal level information are possible. A common way to examine individual-level data is qualitative content analysis, which aims to identify manifest content through category formation. Analogous to usual approaches (e.g., Mayring, 2014), CAMs can be coded inductively and deductively; that is, researchers look at the individual CAMs and rate them according to specific categories. In the earlier example study by Reuter et al. (2021), the pre- and post- CAMs were also examined qualitatively by having two raters independently form categories to rate the within-subjects CAM changes; example categories are *positive development* or *similar to before*. The ratings of each CAM were summarized separately for the experimental and control group, and then compared.

Another option is to focus more on the semantic content of the CAMs, with a thematic analysis. This involves examining which key themes or aspects emerge regarding the research topic. Such a thematic analysis can be implemented, for example, by automatically creating a list of all concepts mentioned in the CAMs and then categorizing these concepts. An example approach can also be found in Reuter et al. (2021), where the authors first performed an automated categorization (e.g., using word databases) and then manually completed the remaining categorization to create a frequency list for the most emerging topics. However, as the tabular data include the information recorded in each node, the process can be implemented using existing research tools like LegislatoR, Lexicoder, Linguistic Inquiry and Word Count (LIWC), or Python. Finally, this combination of qualitative and quantitative evaluations allows CAMs to be used as part of method triangulation or a mixed-methods approach.

## 7. Summary

We introduced the method of cognitive-affective mapping as a tool to assess cognitive and affective mental representations as a network. The method was first presented by Thagard (2010) as an extension of his theory of *emotional coherence*, and in subsequent years, it was used primarily by researchers to restructure and visualize existing data by drawing CAMs. Recently, the release of the new user-friendly software tool called *Valence* has extended the applicability of CAMs as a research tool, allowing individuals to draw CAMs to visualize their experience and impressions on a given topic in a cognitive-affective network.

This development opens up new study design options, such as experimental approaches, but also new possibilities for network analysis. Because of larger data sets, classical network parameters from graph theory, such as density or centrality, gain importance. In addition, the original qualitative analysis options are still available, but with added analytic potential attributable to the updated data exporting capability. The method thus bridges gaps between quantitative and qualitative paradigms and combines advantages from both worlds: CAMs are vivid and easily understandable, while at the same time capable of representing complex relationships. Furthermore, a free response format is possible, freeing participants from the restrictions of traditional closed survey measures, while at the same time generating large amounts of accessible and analyzable data. The CAMs are thus versatile and a promising tool to access and visualize human experience.

## Acknowledgments

This study was funded by the Deutsche Forschungsgemeinschaft (DFG, German Research Foundation) under Germany's Excellence Strategy - EXC-2193/1 - 390951807.